\documentstyle[bbm,amsfonts,epsfig,12pt]{article}

\renewcommand{\a}{\alpha}

\newcommand {\eqref} [1] {(\ref {#1})}
\newcommand {\slsh} [1] {\not{\hbox{\kern-2pt${#1}$}}}

\newcommand {\beq} {\begin{equation}}
\newcommand {\eeq} {\end{equation}}
 \newcommand {\ber}{\begin{eqnarray*}}
 \newcommand {\eer} {\end{eqnarray*}}
\newcommand {\bea}{\begin{eqnarray}}
 \newcommand {\eea} {\end{eqnarray}}
\newcommand{\rmd}{{\rm d}}

\def\Acknowledgements{\bigskip  \bigskip {\begin{center} \begin{normalsize}
             \bf ACKNOWLEDGEMENTS \end{normalsize}\end{center}}}

\begin{document}
\begin{titlepage}
\vskip 1cm
\rightline{CERN-TH/2002-116}
\vskip 3cm
\centerline{{\Large \bf Orbiting Strings in AdS Black Holes}}

\vskip 0.5cm

\centerline{{\Large \bf and ${\cal N}=4$ SYM at Finite Temperature}}
\vskip 1cm
\centerline{{\bf A. Armoni},\,\,{\bf J. L. F. Barb\'on}\footnote{On leave
from Departamento de F\'{\i}sica de Part\'{\i}culas da Universidade de
Santiago de Compostela, Spain.} \,and \,{\bf A. C. Petkou}}
\vskip 0.5cm
\centerline{ \it Theory Division, CERN}
\centerline{\it CH-1211 Geneva 23, Switzerland}
\vskip 0.3cm
\centerline{\it adi.armoni@cern.ch, barbon@cern.ch,  tassos.petkou@cern.ch}
\vskip 1cm

\begin{abstract}

Following Gubser, Klebanov and Polyakov [hep-th/0204051], we study
strings in AdS black hole backgrounds.  With respect to the
pure AdS case, rotating strings 
are replaced by orbiting strings. We interpret these orbiting strings
as  CFT states of large spin  similar to glueballs propagating through
a gluon plasma. The energy and the spin of the 
orbiting string configurations 
are associated with the energy and
the spin of states in the dual finite temperature ${\cal N}=4$ SYM
theory. We analyse in particular the limiting cases of short and long
strings. Moreover, we perform a  thermodynamic study of the
 angular momentum transfer
from the glueball to the plasma by considering string orbits around
rotating AdS black holes. We find that standard expectations, such as
the   complete thermal  dissociation of the glueball, are borne out after
subtle properties of rotating AdS black holes are taken into account.

\end{abstract}
\end{titlepage}

\section{Introduction}

\noindent

The AdS/CFT conjecture
 \cite{Maldacena:1997re,Gubser:1998bc,Witten:1998qj}
 is used to make predictions about the strong
 coupling regime of ${\cal N}=4$ super Yang--Mills theory.
 However, most of the analysis done
 so far is restricted to the supergravity limit where the curvature is
 small and $\a' {\cal R}$ corrections can be neglected (see
 \cite{Aharony:1999ti} for a review).

 Various ideas regarding extensions of
 the AdS/CFT 
 correspondence beyond the supergravity limit have been put forward by
 Polyakov \cite{Polyakov:2001af}. In another interesting recent
 development, the
 authors of \cite{Berenstein:2002jq} considered a particular
 limit of large $R$ charge where the sigma model is exactly solvable.
 Shortly afterwards GKP \cite{Gubser:2002tv} have shown that the
 results of \cite{Berenstein:2002jq} can be rederived
 by considering appropriate configurations of
classical rotating strings in ${\bf S}^5$, i.e.
 solitons
of the nonlinear sigma model on $AdS_5 \times {\bf S}^5$. Moreover, a
 rotation in the ${\bf S}^3$ inside the AdS$_5$ part of the metric was used
 to make a prediction about the anomalous dimensions of states as a
 function of their spin in the strong coupling limit, even at large spin !

 The prescription is as follows: by using the Nambu--Goto action, the
 energy $E$ and the spin $S$ of a rotating string inside AdS$_5$ can be
 computed. The energy and spin are then re-interpreted as the
 energy and spin of twist two states in the boundary theory of the form
\beq
\Phi \nabla ^S \Phi |\,0\,\rangle\;.
\eeq
Such an interpretation is natural, since twist two states are dual to
 massive string modes, i.e. they are absent in the supergravity
 spectrum \cite{Arutyunov:2000ku}.\footnote{Higher twist states that are dual
 to massive string modes are also expected to be described by the GKP
 prescription.} Then, due to conformal invariance
the energy on ${\bf S}^3$  can be identified with the anomalous dimension $\Delta$
on ${\bf R}^4$. The
 remarkable result of GKP \cite{Gubser:2002tv},
\beq
\Delta - S \approx \frac {\sqrt \lambda}{\pi} \ln \frac{S}{ \sqrt \lambda} \,,
\label{GKPresult}
\eeq
can be compared with the perturbative result for conformal gauge field
 theories \cite{Floratos:1977au,Korchemsky:1988si,Dolan:2001tt}

\beq
\Delta - S = f( \lambda)  \ln S \,,
\eeq
where $f(\lambda)$ is a power series in $\lambda$. The analysis of GKP was
 afterwards extended beyond the Nambu--Goto action by
 \cite{Frolov:2002av} (see also \cite{Russo:2002sr}),
 confirming the $\ln S$ behavior beyond the $\lambda \rightarrow
 \infty$ limit.

 This  prescription is expected to hold
for more general cases such as non-supersymmetric and non-conformal
 theories. Here we would like to calculate, using the GKP prescription,
the relation between the energy and spin of states dual to string
modes at finite temperature in ${\cal N}=4$ SYM. To
this end we will consider the motion of classical strings in AdS$_5$
black hole background. In this case we cannot identify    the energy with
the scaling dimension, since we are considering a four-dimensional  field
theory on ${\bf S}^3\times {\bf S}^1$.
The novel phenomenon is that physical states
 should be described by orbiting strings {\em outside} the black hole.
We find corrections to the energy as a function of the spin both
for short and long strings. Our main result is that states with large
spin at large temperature, but $(RT)^4 \sqrt \lambda \ll S$, obey the following
relation
\beq \label{result}
ER -S \approx {\sqrt \lambda \over 2\pi} \ln { S\over (RT)^4 \sqrt \lambda }\,.
\eeq
We also study the thermodynamical stability of these strings with
respect to angular momentum leakage to the black hole. We find
that these strings are unstable and eventually decay into a rotating black hole
that harbours most of the initial spin. This leads to a natural interpretation
of the decay process as the ``melting" of a heavy glueball in the
gluon plasma.

The organization of the manuscript is as follows: in section 2 we
review the general prescription of GKP. Section 3 is devoted to the
 case of AdS$_5$ black hole. In section 4 we present our results
 for the specific limits of short and long strings. Finally, in section 5
we propose the gauge-theory interpretation of these results and study
briefly the case of rotating AdS$_5$ black holes.
In section 6 we present our conclusions.

\section{Semiclassical orbiting strings in curved backgrounds}

\noindent

We consider semiclassical string motion in spherically symmetric
5-dimensional spacetimes whose general metric is of the form
\cite{deVega:1996mv,Kar:1997zi}
\bea
& & \label{gmetric}
\rmd s^2 = -f(r)\rmd t^2 +\frac{\rmd r^2}{f(r)} +r^2\rmd\Omega_3^2 \\
& & 
\rmd \Omega _3 ^2 = \rmd \theta ^2 + \sin ^2 \theta \rmd \phi_1^2 + 
\cos ^2 \theta \rmd \phi_2 ^2
\eea
We are interested in the special cases
\bea
\label{ads5}
f(r) =1+\frac{r^2}{R^2} &\rightarrow & \mbox {AdS$_5$ in global
  coordinates}\,,\\
\label{ads5BH}
f(r) = 1+\frac{r^2}{R^2} -\frac{M}{r^2} &\rightarrow& \mbox{AdS$_5$
  black hole}\,.
\eea
The parameter $M$ is related to the black hole mass and for large
  temperatures we have 
 $M/ R^2 \sim (R T)^4$ \cite{Witten:1998zw}. Consider classical string
  configurations extending in the radial 
 $r$ direction of spacetime and rotating along the $\phi$ angle with
  constant angular velocity $\dot{\phi_1} =\dot {\phi}= \omega$, that
  is we consider circular orbits. This is achieved by choosing
\beq
\label{gauge}
X^{\mu}(\tau,\sigma) = (t, r(\sigma), \phi(t), {\pi \over 2},0)\,,
\eeq
and by identifying $\tau$ and $t$. Then, the  Nambu--Goto action
\beq
\label{NG}
I = - {1 \over 2\pi \a '} \int \rmd\tau\rmd\sigma
\sqrt{-\det G_{\mu \nu} \partial _\alpha X^\mu \partial _\beta X^\nu}\,,
\eeq
takes the following form
\beq
I = - {1 \over 2\pi \a '} \int \rmd \tau \int \rmd \sigma
\sqrt{-G_{rr} \left ({dr\over d\sigma} \right )^2 (G_{tt} + G_{\phi \phi} \dot \phi ^2)  }\,.
\eeq
which yields
\beq
\label{NG1}
I = -  {1 \over 2\pi \a '} \int \rmd \tau \int _{r_{min}}^{r_{max}}\rmd r \sqrt{1 -
  \dot{\phi}^2\frac{r^2}{f(r)}}\,.
\eeq
The integration range $r_{min}\leq r\leq r_{max}$ is determined by the
  condition that the square root in \eqref{NG1} is real. For the
  spacetimes \eqref{ads5} and \eqref{ads5BH}, this will be determined
  in terms of the parameters $\omega$ and $M$.

The energy and angular momentum of the string are then given by
\bea
\label{E}
E &=& {2\over 2\pi \a'} \int _{r_{min}}^{r_{max}}\rmd r \frac{1}{\sqrt{1 -
  \omega^2\frac{r^2}{f(r)}}}\,,\\
\label{Sp}
S &=& {2\over 2\pi \a'}  \int _{r_{min}}^{r_{max}}\rmd r
  \frac{\omega\frac{r^2}{f(r)}}{\sqrt{1 -
  \omega^2\frac{r^2}{f(r)}}}\,.
\eea
The factor of two arises because the string is folded. Note that the
  setup in the presence of a horizon, see figure 1, is different from
  the AdS$_5$ case. In the latter, the string is
  folded around the origin and therefore a factor 4 arises in the
  expressions for energy and spin.

Throughout all intermediate calculations we will set the AdS
radius $R=1$. The radius $R$ will be re-introduced in the final expressions.

\section{The AdS$_5$ black hole background}

\noindent

We will be interested in strings in the AdS$_5$ black hole
background. In the presence of the black hole horizon, the strings will
 rotate outside the horizon and not around the point $r=0$ as in the
AdS case i.e. the strings are {\it orbiting} around the black hole.

Plugging the AdS$_5$ black hole metric \eqref{ads5BH} in \eqref{E} and
\eqref{Sp} we obtain the following expressions
\bea
\label{Ebh}
E &=& {1\over \pi \a'} \int _{r_{min}}^{r_{max}}\rmd r
\sqrt{r^4 + r^2 - M \over (1-\omega ^2)r^4 +r^2 -M}\,,
\\
\label{Spbh}
S &=& {1\over \pi \a'}  \int _{r_{min}}^{r_{max}}\rmd r
 {\omega r^4 \over \sqrt{ ((1-\omega ^2)r^4 +r^2 -M)(r^4 + r^2 -M)}}\,.
\eea
Given $\omega^2$ one should guarantee that the expressions \eqref{Ebh}
and \eqref{Spbh} are real. This requirement will set the boundaries
 of the integrals $r_{min}$ and $r_{max}$.
Indeed by finding the roots of the quadratic equation
\beq
(1-\omega ^2)r^4 + r^2 -M=0 \label{roots}\,,
\eeq
one obtains
\beq
\label{V2}
r_\pm^2 = \frac{1}{2(\omega^2-1)}
\left[1\pm\sqrt{1-4M(\omega^2-1)}\right]\,,
\eeq
such that $r_{min}=r_-$ and $r_{max}=r_+$. In order to have real
roots, we must constraint  the possible values of $\omega ^2$ as
\beq
\omega ^2 -1 \le {1\over 4M} \,.
\eeq
A different way to find the physical condition \eqref{roots} is to
write the conformal constraints $T_{++}=T_{--}=0$ which give
\beq
G_{tt} + G_{\phi \phi} \omega ^2 + G_{rr} \left ( {dr \over d\sigma}
\right ) ^2 = 0\,,
\eeq
yielding
\beq
\label{eom}
 \left ( {dr \over d\sigma}
\right ) ^2 = {( r^4(1-\omega ^2) + r^2  -M )(r^4 + r^2 - M) \over  r^4}\,.
\eeq
The positivity of the rhs of \eqref{eom} yields the physical
constraints on the location of the string. In particular, at
$r_{min},r_{max}$ the derivative with respect to $\sigma$ vanishes
which gives \eqref{roots}.

In addition it is useful to write the roots of the equation
\beq
r^4 + r^2 -M=0\,,
\eeq
as
\beq
r_{H}^2 = {1\over 2} (\sqrt{1+4M}-1) \label{horizon}\,,\,\,\,\,
-a^2 =-{1\over 2} (\sqrt{1+4M}+1)\,,
\eeq
where $r_{H}$ is the horizon. Note that $r_- > r_{H}$, such that the
string orbits outside the black hole. The expressions for the energy and the spin can be written in the
 following form
\bea
\label{Ebh1}
E &=& {1\over \pi \a'} \int _{r_-}^{r_+}\rmd r
\sqrt{(r^2-r_{H}^2)(r^2+a^2) \over (\omega ^2-1)(r^2-r_-^2)(r_+^2-r^2)}\,,
\\
\label{Spbh1}
S &=& {1\over \pi \a'}  \int _{r_{-}}^{r_{+}}\rmd r
 {\omega r^4 \over \sqrt{(\omega
 ^2-1)(r^2-r_-^2)(r_+^2-r^2)(r^2-r_{H}^2)(r^2+a^2)}}\,.
\eea
The physical situation is described in figure 1 below.
\begin{figure}
  \begin{center}
\mbox{\kern-0.5cm
\epsfig{file=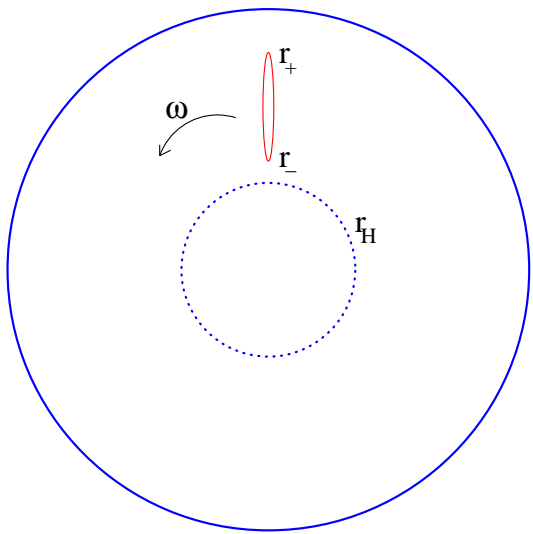,width=7.0true cm,angle=0}}
\label{BHfig}
  \end{center}
\caption{Spinning string in AdS$_5$ black hole background. The string stretches
  outside the black hole. }
\end{figure}
We give the explicit expressions for the integrals in the Appendix. In
the next section we will discuss some specific cases.
\section{Explicit solutions}

\subsection{Short strings}

\noindent

When $\omega ^2 \rightarrow \omega _c ^2 = 1 + {1\over 4M} $ (the
critical value) 
the string shrinks to a
point. In contrast to the pure AdS$_5$, in the AdS$_5$ black-hole case
this solution corresponds to a {\em particle} which rotates outside the
 horizon. The location of the orbit is at $r_+^2=r_-^2= 2M$.

The expressions for the energy and spin in this limit are given by
\bea
& & E = {1\over \pi \a '} \sqrt{4M^2+M \over 2}
\int _{r_-} ^{r_+} {dr \over \sqrt {(r_+-r)(r-r_-)}}=
 {1\over \a '} \sqrt{4M^2+M \over 2}\,, \\
& & S = {1\over \pi \a '} \sqrt{2M^2}
\int _{r_-} ^{r_+} {dr \over \sqrt {(r_+-r)(r-r_-)}}=
 {1\over \a '} \sqrt{2M^2}\,. \label{spinshort}
\eea

The relation between the energy and the spin, expressed in terms
of the spin, becomes temperature independent
\beq
 E R = S \sqrt {1 + {\sqrt \lambda \over 2\sqrt 2 S}}\,.
\eeq
For low temperatures, the large AdS$_5$ black-hole solution does not
dominate over the 
thermal ensemble
 \cite{Witten:1998zw} and therefore the above expression is not valid for
small $S$ (note that from \eqref{spinshort} $S\sim T^4$). However, for large $S$ there should be states
in the spectrum with the above relation between energy and spin.
Namely, given a temperature $T$, there should exist a state with spin
$S\sim T^4$ and energy $E$ such that their difference is a constant
\beq
ER - S \approx \sqrt \lambda \,.
\eeq
\subsection{Long strings}

\noindent

We would like to discuss now the interesting case of long strings.
Long strings are achieved by taking $\omega ^2 \rightarrow 1$. We will
parametrise the limit as
\beq
\omega ^2 = 1 + \eta, \,\,\,\, \eta \rightarrow 0\,.
\eeq
In addition, we wish to take the high temperature limit
\beq
{1\over 4M} = \epsilon, \,\,\,\, \epsilon \rightarrow 0\,,
\eeq
and since $\omega ^2 \le 1 + {1\over 4M}$, we also choose
\beq
{\eta \over \epsilon} \ll 1 \,,
\eeq
such that both the spin and the temperature are large, but the
ratio of the temperature over the spin is small.
The relevant parameters $r_+,r_-,a,r_H$ take the following values in
this limit
\beq
r_+^2 = {1\over \eta},\,\,\, r_-^2 = {1\over 4\epsilon},\,\,\, a^2 =
{1\over 2\sqrt \epsilon}+{1\over 2},\,\,\, r_H^2 = {1\over 2\sqrt
  \epsilon}-{1\over 2}\,.
\eeq
The explicit expression for the energy is
\beq
\label{Elong}
E = {1\over 2\pi \a'} \int _{\sqrt{1\over 4\epsilon}} ^{\sqrt{1\over
    \eta}} \sqrt { r^4 + r^2 \over \eta (r^2 - {1\over
    4\epsilon})({1\over \eta} - r^2)} = {1\over 2\pi \a'} {1\over \eta}
 - {1\over 4\pi \a'} \ln {\eta \over 4\epsilon} \,,
\eeq
and for the spin
\beq
\label{Slong}
S = {1\over 2\pi \a'} \int _{\sqrt{1\over 4\epsilon}} ^{\sqrt{1\over
    \eta}} {\omega r^4 \over \sqrt { \eta (r^2 - {1\over
    4\epsilon})({1\over \eta} - r^2) (r^4 + r^2)}}
 = {1\over 2\pi \a'} {1\over \eta} +
 {1\over 4\pi \a'} \ln {\eta \over 4\epsilon} \,.
\eeq
The equations \eqref{Elong}, \eqref{Slong} lead to the following
 relation between the energy and the spin
\beq \label{lstrings}
ER -S \approx   {\sqrt \lambda \over 2\pi} \ln { S\over (RT)^4 \sqrt
    \lambda }\,. 
\eeq
This is our prediction for the energy of states with large spin at
 large temperature in the strong coupling limit. It is interesting
 to compare the above result \eqref{lstrings} to the finite temperature
 correction to the Wilson loop \cite{Rey:1998bq,Brandhuber:1998bs}.
 In both cases the
 correction to the solitonic state (open strings attached to the
 boundary in the Wilson loop case, or closed orbiting strings in the present
 case) involves a fourth power of the temperature.

 Note also the
 similarity of \eqref{lstrings} to the GKP result \eqref{GKPresult}
 for long strings at zero temperature and the difference by a
 factor of half in front of the logarithmic correction. This
 factor arises naturally since the present 'planetoid' can be viewed as
half of the GKP string after breaking into two pieces at the center
 of the $AdS$ due to the formation of a black hole. Notice however
 that for fixed spin the energy is minimized by a single planetoid.

\section{CFT interpretation}

\noindent

The large AdS$_5$ black hole has a standard holographic interpretation
as the canonical thermal ensemble of the CFT$_4$ on ${\bf S}^3$
 at temperatures well above the critical
temperature $T \gg 1/R$ \cite{Witten:1998zw}. At these temperatures
the thermodynamic 
functions are well approximated by a ``gluon plasma" of $O(N^2)$
particle degrees of freedom. For example, the free energy scales
as $F(T)/T \sim -N^2 \,(R\,T)^3$. On the other hand, the sigma-model
solitons representing the rotating strings have no significant
degeneracy. They are associated to a
single   gauge-invariant  state $|E,S\,\rangle$
 in the Hilbert space of the CFT on
${\bf S}^3$ of energy $E$ and spin $S$. For $S\gg (R\,T)^4 \sqrt
\lambda $ we have
a very long string with $E\approx S$ up to logarithmic corrections,
just as in the zero-temperature theory. Hence, in the regime
$S\gg (R\,T)^4 \sqrt \lambda \gg 1$
we interpret the ``planetoid string" orbiting the black hole as a large-$N$
approximation to the
CFT state
\beq \label{mixx}
 \rho(T,E,S) = \rho_T \;\otimes\;\big(\,|E,S\rangle \,\langle E,S|\,\big)
\;,\eeq
where
\beq
\rho_T = {1\over Z(T)}\;e^{-H/T}, \qquad Z(T) \equiv {\rm
  Tr}\,(\,\rho_T \,)\,, 
\eeq
 is the  density matrix of the canonical ensemble  at temperature $T$.
The direct product representation \eqref{mixx} is less and less accurate
as $S$ approaches $(RT)^4 \sqrt \lambda$. In fact, when $S\sim
(RT)^4\sqrt \lambda$ the
 string degenerates to a point orbiting just above the horizon, and for
smaller values of the spin there are no stable circular orbits that
would represent stationary states. According
to the UV/IR connection, this means that the typical length
 scales of the thermal
ensemble and the gauge-invariant ``glueball" state are well separated
only for $S\gg (RT)^4 \sqrt \lambda $. In this regime we have a very heavy glueball
propagating in a thermal background. Since the glueball energy is
much larger than the thermal energy density, the glueball propagates
without much distortion, to leading order in the large-$N$ expansion.  On
the other hand, for $S\sim (RT)^4 \sqrt \lambda $  we have $E\sim M$ and
the glueball is considerably distorted by the thermal bath (it has
degenerated to a pointlike object in AdS). Finally, for lower spins
the glueball ``melts" in the thermal bath and the corresponding string
enters a spiral orbit and falls behind the black hole horizon.

This is a rather natural physical picture of the fate of heavy particles
interacting with a background plasma. In fact, the apparent stability of
the circular string orbit around the black hole should be an artifact of the
large-$N$ limit. Effects of relative order $O(1/N^2)$ should render the
glueball unstable. This corresponds to interactions mediated by Feynman
diagrams
of toroidal topology in the CFT, or one-loop effects on the gravity side.
At this order we have to consider for example the thermal Hawking radiation
in equilibrium with the black hole. It is clear that both ``gravitational
bremsstrahlung" and  the friction with the
radiation will eventually damp the orbit of the ``planetoid" and transfer
angular momentum to the radiation. Microscopically, this process involves
radiating   small closed strings that fragment out of the lower endpoint
of the long string (the local temperature is higher at the lower endpoint).
 For entropic reasons,
 a significant fraction of
this dissipated angular momentum will end up in the black hole when the
system relaxes to a stationary state.  The precise mechanism of the
decay will be complicated by the possibility of the long string fragmenting
into large pieces first, with subsequent fragmentation into smaller
ones or slow emission of short closed strings.

Thus, we conclude that the orbital strings considered in this paper
are  weakly unstable (within the $1/N$ expansion) towards  falling into
the black hole with the corresponding transfer of rotational energy. On
the CFT side this process corresponds to the gradual melting of a
heavy glueball by interactions with the thermal plasma. Eventually
the glueball cannot be distinguished from thermal fluctuations and
 all the initial angular momentum is transferred to the plasma. The final
stationary state is dual to a rotating AdS black hole such as those
studied in \cite{Hawking:1998kw}.\footnote{Of course,
the thermal radiation in equilibrium with the black hole also carries
part of the angular momentum, but its rotational energy is suppressed
by two powers of $N$ with respect to the black hole.}

In the next subsection we test this scenario by considering the
orbits of stretched strings around rotating AdS black holes.

\subsection{Strings in rotating AdS$_5$ black hole backgrounds and
    thermodynamical stability}

\noindent

Let us consider a thermal ensemble on ${\bf S}^3$ with grand canonical
free energy
$G(T,\Omega)$, where $\Omega$ is the chemical potential associated
to the spin conserved quantum number $S$. It represents the average
angular velocity of the thermal plasma around some  axis in ${\bf S}^3$.
If we work with the canonical  free energy $F(T,S)$,
 evaluated as a function of the total spin  we
have
\beq \label{omegas}
\Omega = {\partial \,F(T,S) \over \partial \,S} \,\,.
\eeq
Subluminal rotation of the thermal ensemble requires $\Omega < 1$ (in units
where $R=1$). In the dual AdS description, $\Omega$ is the angular
velocity of a rotating AdS black hole (in a suitable reference frame).

Equilibrium with respect to angular momentum transfer between the
rotating black hole and the rotating string is achieved when both
chemical potentials are equal. The string ``chemical potential" is
nothing but its angular velocity
\beq \label{angv}
\omega = {\partial\,E(S) \over \partial\,S}\,.
\eeq
For long strings and slowly rotating black holes, we may approximate
$E(S)$ by \eqref{lstrings} and we obtain
\beq \label{superl}
\omega \approx 1+{\sqrt{\lambda} \over 2\pi S} \,.
\eeq
This leads to the somewhat counterintuitive result that the long strings
correspond to glueballs that rotate ``faster than light" around ${\bf
  S}^3$.\footnote{This property 
is  also true for the zero-temperature theory.  Since $\omega$ is
some sort of ``average" angular velocity in the CFT interpretation,
the superluminal motion on the boundary is presumably compatible with
causality. It would be interesting to elucidate this point further in
the light of the UV/IR connection.}   At any rate, the fact that
$\omega >1 > \Omega$ at very large $S$
 means that the orbiting string is always unstable to spin
transfer towards the static black hole.  The  stability condition
\beq
\omega =\Omega\,,
\eeq
would require that eventually we reach a state where
 either $\Omega >1$ or $\omega <1$. The first option is actually excluded
for stable black holes.

The purpose of this subsection is to investigate this question
 by studying equatorial circular orbits of strings around rotating
 AdS black holes. We consider a single common
rotation axis and ``parallel" rotations of both the string and the black hole.
These constraints are natural for a process where the black hole angular
momentum is entirely due to the transfer from the planetoid.

We consider rotating strings in the background of a Kerr
AdS$_5$ black hole with a single rotation parameter $a_1=a$. The metric is
given e.g. in Eq. 2.1 of \cite{Hawking:1999dp}.
Since the
rotation of the string takes place along one of the circles of the
AdS$_5$ black hole background, we only need to consider the above Kerr
metric at $\theta =\pi/2$ and $\phi_1 = \phi, \,\phi_2=0$.
The resulting metric can
then be put in the form
\bea
\label{Kerr1}
\rmd s^2 &=& -\frac{1}{r^2}(\Delta -a^2)\rmd t^2 +\frac{r^2}{\Delta}\rmd
r^2 +\frac{1}{r^2(1-a^2)^2}[ (r^2+a^2)^2 -a^2 \Delta]\rmd
\phi^2\nonumber \\
&& +\frac{2a}{r^2(1-a^2)}[\Delta -(r^2+a^2)]\rmd t\rmd\phi\,, \\
\label{DKerr}
\Delta &=& (r^2+a^2)(1+r^2)-M\,,
\eea
where again we set the AdS radius $R=1$.

There is a subtlety in the holographic map between this metric and
the boundary CFT. In the given coordinates, the asymptotic sphere at
infinity in (\ref{Kerr1}) rotates with angular velocity $\Omega_\infty =-a$.
In order to map the horizon angular velocity to the chemical potential
$\Omega$ in the dual CFT it is convenient to subtract this inertial dragging
of the boundary by redefining the angular coordinate
\beq \label{redphi}
\varphi = \phi + a\,t\,.
\eeq
In these new coordinates the asymptotic boundary of the
 Kerr AdS spacetime is static and the angular
 velocity of the
 black hole is
\beq
\label{KerrO}
\Omega = \Omega_H - \Omega_\infty =  \frac{a(1+{r}_H^2)}{{r}_H^2 +a^2}\,,
\eeq
with the horizon radius ${r}_H$ given by the largest solution of the
 equation
\beq
\label{horKerr}
({r}_H^2 +a^2)(1+{r}_H^2)=M
\,.\eeq

In \cite{Hawking:1999dp} it was
shown that the black hole free energy diverges as $\Omega \rightarrow 1$
from below
\beq
F/T \sim -{N^2 \over 1-\Omega^2} \, (R\,T)^3\;.
\eeq
 Hence, the strongly-coupled CFT can only  tolerate chemical
potentials in the range $\Omega^2 <1$: it takes infinite free energy
to raise the angular velocity to the speed of light on the boundary
${\bf S}^3$.

It is interesting to note that one can have $|\Omega| >1$ for black holes with
negative specific heat. The situation regarding the angular velocity in
the boundary theory is analogous to the occurrence
of $\omega >1$ for rotating strings that eventually become unstable.

The  metric (\ref{Kerr1}) is of the general
axisymmetric form
\beq
\label{axiss}
\rmd s^2 =G_{tt}\rmd t^2 +G_{rr}\rmd r^2 +G_{\phi\phi}\rmd \phi^2
+2\,G_{t\phi} \rmd t \rmd \phi
\,.\eeq
Using the gauge (\ref{gauge}) we can also easily derive integral formulas
for the energy,
angular momentum and action of strings in (\ref{axiss}) as
\bea
\label{Eaxiss}
E &=& \frac{1}{\pi\a'}\int_{r_{min}}^{r_{\max}}\rmd r
\frac{-G_{rr}\left(G_{tt}+\omega G_{t\phi}\right)}{\sqrt{V
    }}\,, \\
\label{Spaxiss}
S &=& \frac{1}{\pi\a'}\int_{r_{min}}^{r_{\max}}\rmd r
\frac{G_{rr}\left(\omega G_{\phi\phi}+G_{t\phi}\right)}{\sqrt{V
    }}\,, \\
\label{Iaxiss}
I &=& -{2\over \omega \alpha'} \int_{r_{min}}^{r_{max}} \rmd r \sqrt{ V}\;.
\eea
The effective potential $V$ is given by
\beq \label{effV}
V\equiv -G_{rr} \,\left(G_{tt} + 2\omega G_{t\phi} + \omega^2 \,G_{\phi\phi}
\right)\,.
\eeq
From the integral representations one can prove the relations
\beq
-{\omega \,I \over 2\pi} = E-\omega\,S, \qquad S=-{\partial \over \partial
\omega} \,\left(-{\omega\,I \over 2\pi} \right)\,.
\eeq
This means that $-\omega I(\omega) /2\pi$ is the Legendre transform
$E(S)$ of the
energy  as a function of the spin. Hence, one has the desired
relation between the angular velocity of the string and the ``spin dispersion
relation"
\beq
\omega = {\partial E(S) \over \partial S}\;.
\eeq

Explicitly, for the Kerr metric (\ref{Kerr1})  we find the effective potential
\beq
\label{V}
V = \frac{1-(\omega+a)^2}{\Delta(1-a^2)}\left[ r^4
   +\left(\frac{1-a^2(\omega +a)^2}{1-(\omega +a)^2}\right) r^2
   -\left(\frac{(
       1-a(\omega+a))^2}{1-(\omega+a)^2}\right)\frac{M}{1-a^2}\right]
\;.
\eeq
From (\ref{V}) we see that it is natural to redefine the
angular velocity for the string as
\beq
\label{omegaeff}
 \omega +a \longrightarrow \omega
\,.\eeq
This is the angular velocity of the string in the coordinate system
 corresponding to the change of variables $\phi =\varphi -a\,t$, exactly
the same as in (\ref{redphi}). Hence, it is the shifted angular velocity
as in (\ref{omegaeff}),
 defined by $\varphi(t)=\omega\,t$, the one  that must  be compared directly
to the thermal chemical potential $\Omega$.

  Given the fact that
$\Omega <1$ for the stable black holes that we are considering, the
stability question reduces to whether
 a non-vanishing rotation parameter of the black hole $a>0$ can induce
finite-energy string orbits with $\omega <1$.

In order to answer this question, we calculate the turning points
$r_{min} = r_-$ and $r_{max} = r_+$ from the equation $V=0$. The result
is
\beq \label{tunn}
r_\pm^2 = {1-a^2 \omega^2 \over 2(\omega^2 -1)} \;\left[\;1\pm\sqrt{
1-{4M (\omega^2 -1) \over (1-a^2)\,(1+a\omega)^2}}\;\right]\;.
\eeq
This expression shows that at soon as $\omega^2 <1$ the outer
endpoint $r_+$ becomes imaginary for any value of $a^2 \leq 1$. This
means that we must necessarily have $\omega^2 >1$ and no stability
is possible between the planetoid and the rotating black hole.

It is interesting to obtain the rough features of the orbits. For
$\omega >1$ the condition that $r_\pm^2 >0$ and real is that
$\omega < 1/a$ and that the argument of the square root in (\ref{tunn})
be positive. The second condition defines a maximal angular velocity
$\omega_c$ that can be explicitly calculated as a function of $a$ and
$M$. For large black hole masses (large temperature) one obtains
\beq \label{axxi}
\omega_c = 1 + (1-a) \,\left[ {(1+a)^3 \over 8M} + \dots\right]\;.
\eeq
The overall factor of $(1-a)$ arises from the fact that $a=1$ requires
$\omega_c =1$.

From the expression (\ref{axxi}) we can deduce that the condition
$\omega < 1/a$ is redundant. This condition can only be relevant for
$a$ close to unity. Let $a=1-\epsilon$, $1/a \approx 1+\epsilon$. Then
for large mass one has
\beq
\label{omegacr1}
\omega_c \approx 1 + {\epsilon \over M} < 1+\epsilon \approx 1/a\,,
\eeq
and we see that the relevant interval of angular velocities is
$1<\omega < \omega_c$.

The rough features of the planetoids are similar to those at $a=0$. As
$\omega \rightarrow \omega_c$ from below the string degenerates to
a point orbiting at
\beq
r^2_+ = r^2_- = 2M {(1-a\omega_c)^2 \over (1-a^2)^2}\;.
\eeq
For large $M$, this becomes by virtue of (\ref{axxi}) 
\beq
r^2_+ = r^2_- \rightarrow  {2M \over (1+a)^2}\;.
\eeq
On the other hand, for $\omega \rightarrow 1$ from above the upper
endpoint tends to infinity $r_+ \rightarrow \infty$ and the lower endpoint
descends to the minimum value
\beq
\label{rminKerr}
r^2_- \rightarrow {M \over (1+a)^2}\;.
\eeq
For $M\gg 1$ the horizon sits at $r^2_H \sim \sqrt{M}$ and the planetoid
is well above the black hole in any of these limits. 

Our main result of this section is that $\Omega < 1<\omega$ for
finite-energy orbiting strings and stable black holes with positive
specific heat. This means that no equilibrium configuration exists
with respect to angular-momentum balance and the planetoid will eventually
fall behind the rotating black hole horizon. It would be interesting to
study the fate of string orbits around ``small" AdS black holes with
negative specific heat. In this case one can have $\Omega >1$ so that
stable orbits are possible, although the black hole itself is unstable.

We have also assumed in our analysis that the orbits were ``direct", i.e.
the black hole and the planetoid rotate in the same sense. One can
also consider ``retrograde" orbits satisfying $\omega\,\Omega <0$. In this
case the final black-hole spin is not entirely due to the transfer from
the planetoid.

\section{Conclusions}

\noindent

We have studied finite temperature effects on the physics of certain
 large spin
states of ${\cal N}=4$ SYM on a three sphere. In the formalism of GKP
 \cite{Gubser:2002tv} these states correspond to folded closed strings
 rotating in 
AdS space. The finite temperature generalization amounts to the study
of similar string configurations in the background of AdS black holes.
The main novelty is that  the black hole  turns the rotating strings
into orbiting strings or ``planetoids" \cite{deVega:1996mv}.

These planetoids stay clear from the black hole horizon and represent
thermally perturbed gauge-invariant states of large energy satisfying
$E\sim S$ with logarithmic accuracy.
Since they correspond to closed-string states we term them ``glueballs".
When the glueball mass is of the order of the thermal energy per gauge
degree of freedom we expect
that the glueball should melt or dissociate in the plasma. This physics is
reflected in the gravitational dual by the degeneration of
the stringy planetoid to a pointlike object rotating not far
 above the horizon. This configuration represents the circular orbit of lowest
spin for a given temperature. All lower orbits decay into the black hole.
 Nevertheless, for
``long string'' configurations, the  corresponding states have energy and
spin much larger than the black hole mass and could represent
quasi-stable glueball states propagating in the thermal bath. For such
states our strong coupling prediction for their energy and spin is
(\ref{result}). 

In subleading orders in the $1/N$ expansion, interaction with the
Hawking thermal radiation should destabilise any orbit, leading to
the progressive melting of the fast glueball into the plasma. We have
checked this idea in the gravity dual by studying orbiting strings
around rotating AdS black holes. We find that the transfer of angular
momentum from the planetoid to the black hole   is always thermodynamically
favoured. Hence, the heavy glueball eventually  melts and transfers all its
angular momentum to the thermal bath.

It would be interesting to study in more detail the fate of the low temperature
GKP soliton across the Hawking-Page phase transition. In particular
the dynamics of string fragmentation versus black hole formation.

These systems provide a very interesting example of the AdS/CFT correspondence,
where one can describe the interaction between
 very special states with peculiar
properties and typical (featureless) states  such as a thermal density
matrix. Once again, a somewhat exotic gravitational dynamics on
the AdS side encodes very nontrivial, yet physically reasonable effects
on the CFT side.

\Acknowledgements

We would like to thank  E. Gardi and S. Wadia for illuminating discussions.
We also thank A. Tseytlin for the critical reading of an earlier
version of the manuscript.

\section*{Appendix}

\noindent

We start with the evaluation of (\ref{Ebh1}). Setting $r^2 =x$ we
obtain
\beq
\label{Ebh2}
E=\frac{1}{2\pi \a' \sqrt{\omega^2-1}}\int_{x_-}^{x_+}\rmd x\,
  \sqrt{\frac{(x-x_H)(x+A)}{x(x-x_-)(x_+ -x)}} \,, x_{\pm} =
  r^2_{\pm}\,, x_H =r^2_H \,, A=a^2
\eeq
Then we set $\rho = x-x_-$ to obtain
\beq
\label{Ebh3}
E = \frac{1}{2\pi \a' \sqrt{\omega^2-1}}\int_0^{\rho_+} \rmd \rho
\sqrt{\frac{(\rho +x_- -x_H)(\rho +A+x_-)}{\rho(\rho+x_-)(\rho_+
    -\rho)}} \,,\,\,\,\,\rho_+ = r_+^2-r_-^2
\eeq
Finally, we set $t=\rho/\rho_+$ to get
\beq
\label{Ebh4}
E = \frac{\sqrt{\rho_+}}{2\pi \a' \sqrt{\omega^2-1}}\int_0^{1} \rmd t
\sqrt{\frac{\left(t +\frac{x_- -x_H}{\rho_+}\right) \left(t
      +\frac{A+x_-}{\rho_+}
    \right)}{t(1-t)\left(t+\frac{x_-}{\rho_+}\right)}}
\eeq
Similar calculations lead to the following expression for the spin
(\ref{Spbh1})
\beq
\label{Spbh2}
S = \frac{\omega\sqrt{\rho_+}}{2\pi \a' \sqrt{\omega^2-1}}\int_0^{1} \rmd t
\frac{\left(t+\frac{x_-}{\rho_+}\right)^{\frac{3}{2}}} {\sqrt{t(1-t)\left(t +\frac{x_- -x_H}{\rho_+}\right) \left(t
      +\frac{A+x_-}{\rho_+}
    \right)}}
\eeq
Clearly, the above two expressions may be considered as representing
generalized
hypergeometric series in the three variables $x_-/\rho_+$, $(x_-
-x_H)/\rho_+$ and $(A+x_-)/\rho_+$. Nevertheless, in the interesting
cases of long and short strings the expressions simplify.

{\bf a)} Short strings. In this case $\omega\rightarrow \omega_{c}$
and $\rho_+\rightarrow 0$ therefore
the relevant representations of (\ref{Ebh3}) and (\ref{Spbh2}) are
\bea
\label{Eshort}
E &=& \frac{1}{2\pi \a' \sqrt{\omega^2-1}}\sqrt{\frac{(x_--x_H)(A+x_-)}{x_-}}
\int_0^1\rmd t
\sqrt{\frac{\left(
  1+t\frac{\rho_+}{x_--x_H}\right)
\left(1+t\frac{\rho_+}{A+x_-}\right)}{
t(1-t)\left(1+t\frac{\rho_+}{x_-}\right)}}
\\
\nonumber
S &=&
\frac{\omega}{2\pi \a' \sqrt{\omega^2-1}}
\frac{(x_-)^{\frac{3}{2}}}{\sqrt{(x_--x_H)(A+x_-)}}
\int_0^1\rmd t
\sqrt{\frac{\left(1+t\frac{\rho_+}{x_-}\right)^3}{t(1-t)\left(
  1+t\frac{\rho_+}{x_--x_H}\right)
\left(1+t\frac{\rho_+}{A+x_-}\right)}}
\\
 \label{Spshort}
\eea
Then, given the fact that
\beq
\label{omegacr}
\omega_{c}^2 -1 =\frac{1}{4M}
\eeq
we find
\beq
\label{xx}
(x_- -x_H)(A+x_-) = 4M^2\left(1+\frac{1}{4M}\right)
\eeq
and taking the $\rho _+ \rightarrow 0$ limit of (\ref{Eshort}) and
(\ref{Spshort}) we obtain
\bea
\label{Eshort1}
E &= & \sqrt{2\lambda}M\left(1+\frac{1}{4M}\right)^{1/2}
\\
\label{Spshort1}
S &=& \sqrt{2\lambda} M
\eea
This leads to
\beq
\label{ESpshort}
ER = S\left(1+\frac{\sqrt{\lambda}}{2\sqrt{2}S}\right)^{1/2}
\eeq

{\bf b)} Long strings. In this case we set $\omega^2-1 =\eta$ with
$\eta\rightarrow 0$ and we obtain from the explicit expressions above
\bea
\nonumber
E &=& \frac{\sqrt{\rho_+}}{2\pi \a' \sqrt{\eta}} \int _0^1\rmd t
\sqrt{\frac{\left(t+\frac{x_-}{\rho_+}\right)}{t(1-t)}}
  \left[1+\frac{A-x_H}{\rho_+}\left(t+\frac{x_-}{\rho_+}\right)^{-1}
    -\frac{x_H
      A}{\rho_+^2}\left(t+\frac{x_-}{\rho_+}\right)^{-2}\right]^{\frac{1}{2}}
 \\
\label{Elong1}
\\
\nonumber
S &=& \frac{\omega\sqrt{\rho_+}}{2\pi \a' \sqrt{\eta}} \int _0^1\rmd t
\sqrt{\frac{\left(t+\frac{x_-}{\rho_+}\right)}{t(1-t)}}
  \left[1+\frac{A-x_H}{\rho_+}\left(t+\frac{x_-}{\rho_+}\right)^{-1}
    -\frac{x_H
      A}{\rho_+^2}\left(t+\frac{x_-}{\rho_+}\right)^{-2}\right]^{-\frac{1}{2}}
\\
\label{Splong1}
\eea
From the above we easily see that the leading divergence is the same
for both $E$ and $S$ and equal to
\beq
\label{ESPdiv}
E_0 =S_0 \approx \frac{\sqrt{\lambda}}{\eta}
\eeq
To evaluate the corrections to the leading divergence we consider the
difference $E-S$. From the above it is not difficult to see that the
leading term in that difference is
\beq
\label{ESpdiff}
E -S \approx {1\over 2\a '} \left(1+ M\eta \right)^{-\frac{1}{2}}{}
_2F_1\left(\frac{1}{2},\frac{1}{2};1;\frac{1}{1+ M\eta }\right)
\eeq
In the limit $M\eta \rightarrow 0$ the hypergeometric function
has a logarithmic behavior that leads to
\beq
\label{ESplongfin}
ER - S \approx
\frac{\sqrt{\lambda}}{2\pi}\ln\frac{S}{M\sqrt{\lambda}} \sim
\frac{\sqrt{\lambda}}{2\pi}\ln\frac{S}{(RT)^4\sqrt{\lambda}}.
\eeq


\begin{thebibliography}{99}

\bibitem{Maldacena:1997re}
J.~M.~Maldacena,
``The large $N$ limit of superconformal field theories and supergravity,''
Adv.\ Theor.\ Math.\ Phys.\  {\bf 2}, 231 (1998)
[Int.\ J.\ Theor.\ Phys.\  {\bf 38}, 1113 (1999)]
[arXiv:hep-th/9711200].

\bibitem{Gubser:1998bc}
S.~S.~Gubser, I.~R.~Klebanov and A.~M.~Polyakov,
``Gauge theory correlators from non-critical string theory,''
Phys.\ Lett.\ B {\bf 428}, 105 (1998)
[arXiv:hep-th/9802109].

\bibitem{Witten:1998qj}
E.~Witten,
``Anti-de Sitter space and holography,''
Adv.\ Theor.\ Math.\ Phys.\  {\bf 2}, 253 (1998)
[arXiv:hep-th/9802150].

\bibitem{Aharony:1999ti}
O.~Aharony, S.~S.~Gubser, J.~M.~Maldacena, H.~Ooguri and Y.~Oz,
``Large N field theories, string theory and gravity,''
Phys.\ Rept.\  {\bf 323}, 183 (2000)
[arXiv:hep-th/9905111].

\bibitem{Polyakov:2001af}
A.~M.~Polyakov,
``Gauge fields and space-time,''
arXiv:hep-th/0110196.

\bibitem{Berenstein:2002jq}
D.~Berenstein, J.~M.~Maldacena and H.~Nastase,
``Strings in flat space and pp waves from N = 4 super Yang Mills,''
JHEP {\bf 0204}, 013 (2002)
[arXiv:hep-th/0202021].

\bibitem{Gubser:2002tv}
S.~S.~Gubser, I.~R.~Klebanov and A.~M.~Polyakov,
``A semi-classical limit of the gauge/string correspondence,''
arXiv:hep-th/0204051.

\bibitem{Arutyunov:2000ku}
G.~Arutyunov, S.~Frolov and A.~C.~Petkou,
``Operator product expansion of the lowest weight CPOs in N = 4  SYM(4) at strong coupling,''
Nucl.\ Phys.\ B {\bf 586}, 547 (2000)
[Erratum-ibid.\ B {\bf 609}, 539 (2001)]
[arXiv:hep-th/0005182].

\bibitem{Floratos:1977au}
E.~G.~Floratos, D.~A.~Ross and C.~T.~Sachrajda,
``Higher Order Effects In Asymptotically Free Gauge Theories: The Anomalous Dimensions Of Wilson Operators,''
Nucl.\ Phys.\ B {\bf 129}, 66 (1977)
[Erratum-ibid.\ B {\bf 139}, 545 (1978)].

\bibitem{Korchemsky:1988si}
G.~P.~Korchemsky,
``Asymptotics Of The Altarelli-Parisi-Lipatov Evolution Kernels Of Parton Distributions,''
Mod.\ Phys.\ Lett.\ A {\bf 4}, 1257 (1989).

\bibitem{Dolan:2001tt}
F.~A.~Dolan and H.~Osborn,
``Superconformal symmetry, correlation functions and the operator product  expansion,''
Nucl.\ Phys.\ B {\bf 629}, 3 (2002)
[arXiv:hep-th/0112251].

\bibitem{Frolov:2002av}
S.~Frolov and A.~A.~Tseytlin,
``Semiclassical quantization of rotating superstring in AdS(5) x S**5,''
arXiv:hep-th/0204226.

\bibitem{Russo:2002sr}
J.~G.~Russo,
``Anomalous dimensions in gauge theories from rotating strings in $AdS_5\times S^5$,''
arXiv:hep-th/0205244.

\bibitem{deVega:1996mv}
H.~J.~de Vega and I.~L.~Egusquiza,
``Planetoid String Solutions in 3 + 1 Axisymmetric Spacetimes,''
Phys.\ Rev.\ D {\bf 54}, 7513 (1996)
[arXiv:hep-th/9607056].

\bibitem{Kar:1997zi}
S.~Kar and S.~Mahapatra,
``Planetoid strings: Solutions and perturbations,''
Class.\ Quant.\ Grav.\  {\bf 15}, 1421 (1998)
[arXiv:hep-th/9701173].

\bibitem{Witten:1998zw}
E.~Witten,
``Anti-de Sitter space, thermal phase transition, and confinement in  gauge theories,''
Adv.\ Theor.\ Math.\ Phys.\  {\bf 2}, 505 (1998)
[arXiv:hep-th/9803131].

\bibitem{Rey:1998bq}
S.~J.~Rey, S.~Theisen and J.~T.~Yee,
``Wilson-Polyakov loop at finite temperature in large N gauge theory and  anti-de Sitter supergravity,''
Nucl.\ Phys.\ B {\bf 527}, 171 (1998)
[arXiv:hep-th/9803135].

\bibitem{Brandhuber:1998bs}
A.~Brandhuber, N.~Itzhaki, J.~Sonnenschein and S.~Yankielowicz,
``Wilson loops in the large N limit at finite temperature,''
Phys.\ Lett.\ B {\bf 434}, 36 (1998)
[arXiv:hep-th/9803137].

\bibitem{Hawking:1998kw}
S.~W.~Hawking, C.~J.~Hunter and M.~M.~Taylor-Robinson,
``Rotation and the AdS/CFT correspondence,''
Phys.\ Rev.\ D {\bf 59}, 064005 (1999)
[arXiv:hep-th/9811056].

\bibitem{Hawking:1999dp}
S.~W.~Hawking and H.~S.~Reall,
``Charged and rotating AdS black holes and their CFT duals,''
Phys.\ Rev.\ D {\bf 61} (2000) 024014
[arXiv:hep-th/9908109].





\end{thebibliography}
\end{document}